\documentclass[usegraphicx]{mn2e}

\title{Simultaneous optical and near-IR photometry of 4U 1957+115 - a missing secondary star}
\author[Hakala Pasi, Muhli Panu \& Charles Phil]{Pasi Hakala$^{1}$\thanks{E-mail:
pahakala@utu.fi},
Panu Muhli,$^{2}$
Phil Charles$^{3,4}$\\
$^{1}$Finnish Centre for Astronomy with ESO (FINCA), V\"ais\"al\"antie 20, University of Turku, FIN-21500, Piikki\"o,
Finland.\\
$^{2}$National Land Survey of Finland, PL 84, 00521 Helsinki, Finland.\\
$^{3}$School of Physics and Astronomy, University of Southampton, Highfield, Southampton SO17 1BJ, UK\\
$^{4}$4 Astrophysics, Cosmology and Gravity Centre (ACGC), University of Cape Town, Private Bag X3, Rondebosch 7701, South Africa
 }
\begin{document}

\date{}

\pagerange{\pageref{firstpage}--\pageref{lastpage}} \pubyear{2002}

\maketitle

\label{firstpage}

\begin{abstract}
We report the results of quasi-simultaneous optical and NIR photometry of the low-mass X-ray binary, 4U 1957+115. Our observations cover
B, V, R, I, J, H and K-bands and additional time-series NIR photometry. We measure
a spectral energy distribution, which can be modelled using a standard multi-temperature accretion disc, where the disc temperature and radius follow a power-law relation. Standard accretion disc theory predicts the power law exponent to be -3/4, and this yields,
perhaps surprisingly, acceptable fits to our SED. Given that the source is a persistent X-ray source, it is however likely that 
the accretion disc temperature distribution is produced by X-ray heating, regardless of its radial dependence. Furthermore, we find no evidence for any emission from the secondary star at any wavelength. 
However, adding a secondary component to our model allows us to derive a 99\% lower limit of 14 or 15 kpc based on Monte Carlo simulations and using either an evolved K2 or G2V secondary star respectively. In $>$60\% of cases the distance is $>$80kpc. Such large distances favor models with a massive ($>$15 M$_{\odot}$) black hole primary. Our quasi-simultaneous J and V-band time-series photometry, together with the SED, reveals that the optical/NIR emission must originate in the same region i.e. the accretion disc. The likely extreme mass ratio supports suggestions that the accretion disc
must be precessing which, depending on the length of the precession period, could play a major part in explaining the variety of optical light curve shapes obtained over the last two decades.
\end{abstract}

\begin{keywords}
Accretion -- black holes, accretion discs, X-ray binaries.
\end{keywords}

\section{Introduction}

4U 1957+115 (V1408 Aql) is a persistently bright low mass X-ray binary (LMXB). It was first detected by {\it Uhuru} (Giacconi et al. 1974), and its optical counterpart was later identified by Margon, Thorstensen \& Bowyer (1978). The source also appears in an early list of black hole candidates (White \& Mason 1985), as it shows an ultrasoft X-ray spectrum (White \& Marshall 1984), but currently there is no firm evidence on the nature of the compact object. The source was noted to vary in the optical (Motch et al. 1985), and  Thorstensen (1987) discovered a 9.33-h sinusoidal V-band modulation. This period, if of orbital origin, implies that a Roche-lobe filling, main-sequence donor would have a mass of $\sim$ 0.11$\times P_{orb}$(h) = 1.0 M$_\odot$. Intriguingly, no optical spectroscopic signature of the donor has been found (Shahbaz et al. 1996).

Observations by a variety of X-ray missions have painted a mixed picture for the nature of the compact object. 
Earlier studies (Yaqoob, Ebisawa and Mitsuda 1993, Singh, Apparao and Kraft 1994) favor a neutron star model, whilst more
recent work seems to suggest a black hole primary (Ricci, Israel \& Stella 1995, Wijnands, Miller \& van der Klis 2002, Nowak et al. 2008, 2012). Nowak \& Wilms (1999) compare the merits of both models and suggest that the source probably contains a warped precessing disc, as also suggested by Hakala, Muhli \& Dubus (1999, hereafter HMD99) based on long-term changes in the optical orbital light curve.
Recent optical studies have produced mixed results, with Russell et al. (2010) reporting a non-detection for the optical period, 
and Bayless et al. (2011) claiming a pure sinusoidal modulation due to the X-ray heating of the secondary.  

There is no clear X-ray modulation over the orbital period. However, the X-ray luminosity varies by a factor of $\sim$3 over a time scale of a few months (Nowak \& Wilms 1999). Furthermore, there is some tentative evidence for a 117 d or 250-260 d superorbital period from {\it RXTE ASM} data (Nowak \& Wilms 1999). Recent {\it Chandra, XMM-Newton, RXTE} and {\it Suzaku} observations by Nowak et al. (2008, 2012) suggest that the system contains a rapidly rotating black hole, the mass of which hinges on determining the distance to the system. 

\begin{table}
\centering
\caption{4U 1957+115 J-band time-series observing log}
\begin{tabular}{ccccc}
\hline
\hline
 Start date & exposure (s) & length (d) & phase & filter \\
\hline
2452477.4755  &  210x(10x6) & 0.20 & 0.00-0.51 & J  \\
2452478.4397  &  260x(10x6) & 0.25 & 0.48-1.12 & J  \\
2452479.4712  &  180x(10x6) & 0.23 & 0.13-0.72 & J  \\
 \hline
\hline
\end{tabular}
 \medskip
 
Phase has an arbitrary zero point on the 9.33h period. Observations from the first two nights were used to
construct the SED. Additionally, four V-band points where obtained during the J-band time series observations.
\end{table}

\section{Observations}

We have observed the source with the 2.5m Nordic Optical Telescope (NOT) located at the Roque de los Muchachos observatory, La Palma
using a combination of StanCam and NOTCam on July 21-23, 2002 (details in Table 1). StanCam is a stand-by optical CCD located within the Cassegrain instrument rotator of NOT and consists of a TEK 1024x1024 back-side illuminated thinned CCD with a 3x3' field of view. It is always available and is fed though a 45 degree insertable flat mirror. NOTCam is a NIR camera operated at the Cassegrain focus. It is based on a 1024x1024 Rockwell ``Hawaii«« HgCdTe array and has a field of view of 4x4'. 

These two instruments can be operated in a quasi-simultaneous
mode taking exposures in both optical and NIR bands in a sequence. We have obtained exposures in B,V,R,I,J,H and K-bands together with a time series in J and V-bands (Fig. 1). Both the optical and NIR observations have been reduced using IRAF and standard reduction techniques. A multipoint dithering mode was used in the NIR. Further details of the time-series observations can be found
in Table 1. In order to calibrate the optical and NIR observations, several standard stars were observed. In particular, we used five NIR standards from Hunt et al. (1998) to calibrate the J,H and K magnitudes. The SED, tabulated in Table 2 and plotted in Fig. 2, is based on aperture photometry using the standard ``phot«« task of IRAF.         

\begin{figure}
\includegraphics[scale=0.5]{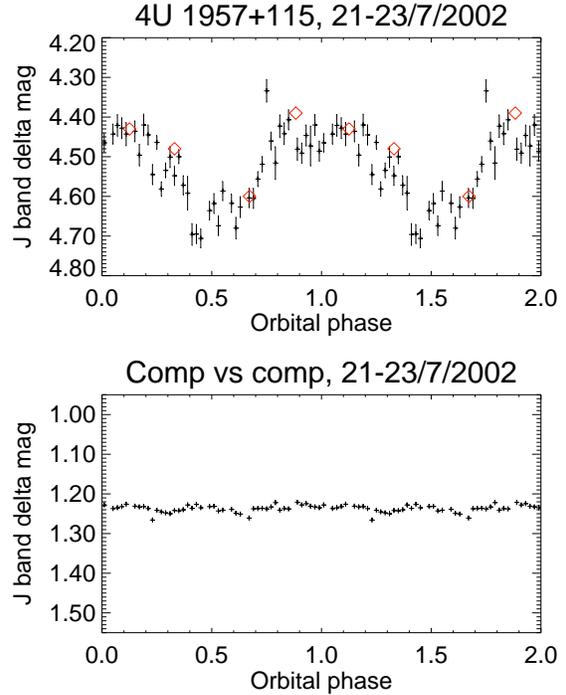}
 \caption{The July 21-23 2002 J-band data folded on the 9.33h orbital period (top). The red diamonds denote quasi-simultaneous V-band
 observations shifted to the same delta mag. range. The lower plot shows the delta magnitude between two comparison stars in the same field.}
\end{figure}

\section{Modelling the spectral energy distribution }

The resulting SED, plotted in Fig 2., shows a very blue continuum. This is in accordance with the low hydrogen column density of 1.0-1.1$\times10^{21}$estimated from the X-ray spectra (Nowak et al. 2008) and thus the interstellar extinction is small (it is still included in our modelling as a free parameter). 

Simple visual examination of the shape of the SED appears to suggest that there might be a slightly different power-law slope
for the optical (B,V,R,I) and the NIR (J,H,K) bands. This would be expected from a cool, low-mass donor. However,
we will show in subsequent modelling that this clearly is not the case.

In order to quantify the level of any contribution by the secondary star, and to investigate the origin of the continuum, we have proceeded to fit it initially with a traditional accretion  
disc model (Shakura \& Sunyaev, 1973), where we assume an optically thick disc with a radial power-law temperature dependance. 
The model consists of annular rings whose temperature varies only as a power-law with radius from the centre.
In order to keep our model simple enough, we have fixed the ratio of inner and outer disc radii to 0.01. We are thus excluding
the inner 0.01\% of the disc area. Selecting a different ratio (eg. 0.001) did not affect the results significantly, since the contribution
from the excluded area is very small. We still retain the same disc temperature profile i.e. the $T_{e}$ at 0.01 radius remains the same.

\begin{table}
  \centering
   \caption{The spectral energy distribution of 4U 1957+115.}
\begin{tabular}{ccc}
\hline
\hline
 Filter & Flux (erg/\AA/cm$^2$/sec) & Magnitude\\
\hline
B   &   16.6 $\pm 0.78 \times 10^{-17}$ & 18.97 $\pm 0.05$ \\
V   &    9.67 $\pm 0.27 \times 10^{-17}$ & 18.95 $\pm 0.03$\\
R   &   6.35 $\pm 0.30 \times 10^{-17}$ & 18.70 $\pm 0.05$\\ 
I     &  3.24 $\pm 0.15 \times 10^{-17}$ & 18.66 $\pm 0.05$\\
J    &  1.47 $\pm 0.10 \times 10^{-17}$ & 18.32 $\pm 0.07$\\ 
H   &  0.64 $\pm 0.061 \times 10^{-17}$ & 18.15 $\pm 0.10$ \\
K   &  0.22 $\pm 0.026 \times 10^{-17}$ & 18.15 $\pm 0.12$\\
\hline
\hline
\end{tabular}
\end{table}

The free parameters are the disc inner radius temperature and the power-law index $\alpha$ that determines the dependence of local disc temperature on the disc radius i.e. $T_{e}(r) \propto r^{\alpha}$ and the amount of interstellar extinction $E_{B-V}$. 

We are unable to obtain a formally good fit to the data using the errors given in Table 2. The values of these typically 
range from 3-5\% in optical to 11\% in K-band. The scatter of our data about the best fit suggests that the formal error is not sufficient and does not contain all the systematics, which are likely to be dominated by the intrinsic flickering/variability of the accreting source,
as demonstrated in both previous and current (e.g. Fig. 1) light curves. Note that the different photometric bands were obtained in a sequence, not simultaneously. We have taken this into account by examining the residuals to the fit and have increased the errors in each band to 15\%. After doing this, we find that we can fit the SED with an accretion disc model, as described above, and a best fit is overplotted in Fig. 2. 

We have determined the errors for the fit parameters and further examined the correlations and the parameter space in general by carrying out a Monte Carlo analysis. This involved creating 20000 synthetic datasets (SEDs) based on a best fit (and the 15\% errors) and fitting these with the model. This exercise revealed that different combinations of fitting parameters are capable of producing
equally good fits, since the inner disc temperature and the radial temperature index $\alpha$ are correlated. This led to
a bimodal distribution of possible models, where we could typically fit the SED either using a low inner disc temperature of the order
of 10000-15000K coupled with a very steep temperature profile and rather unphysical normalisations or an inner disc temperature of 70000-90000K with $\alpha = -0.75$.  The first option is not physical, since it would imply impossibly low disc temperatures for most of the disc. Single black-body temperature fits are also ruled out and the effect of interstellar reddening is found to be negligible.

Secondly, we have experimented by adding a solar-like 5800K (black-body) component to the model to see whether including/forcing  a secondary in the fit would change the temperature distribution of the resulting accretion disc model and in order to derive upper limits for the emission from the secondary. In this case we could fit
the SED also with a uniform temperature disc, but the model for any contribution from the secondary is not physical, since it requires the secondary star projected area to be an order of magnitude larger than that of the whole accretion disc.
   
In conclusion, we can model the SED using an inner disc temperature of 81000K $\pm$ 26000K together with a  conventional radial temperature dependence value for the power-law i.e. $\alpha$=-0.75 and $E_{B-V}$=0.0 $\pm$ 0.07, $\chi^2$=3.92, 2 degrees of freedom (p=0.14). No contribution from the secondary star is required for this fit. 

\begin{figure}
\includegraphics[scale=0.37,angle=90]{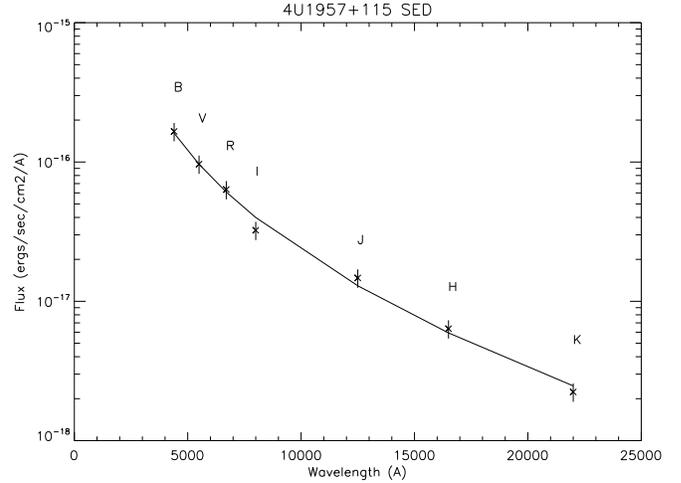}
 \caption{The B,V,R,I,J,H,K spectral energy distribution (SED) of 4U 1957+115, together with our accretion disc model.
See text for details.}
\end{figure}

\begin{figure}
\includegraphics[scale=0.55,angle=0]{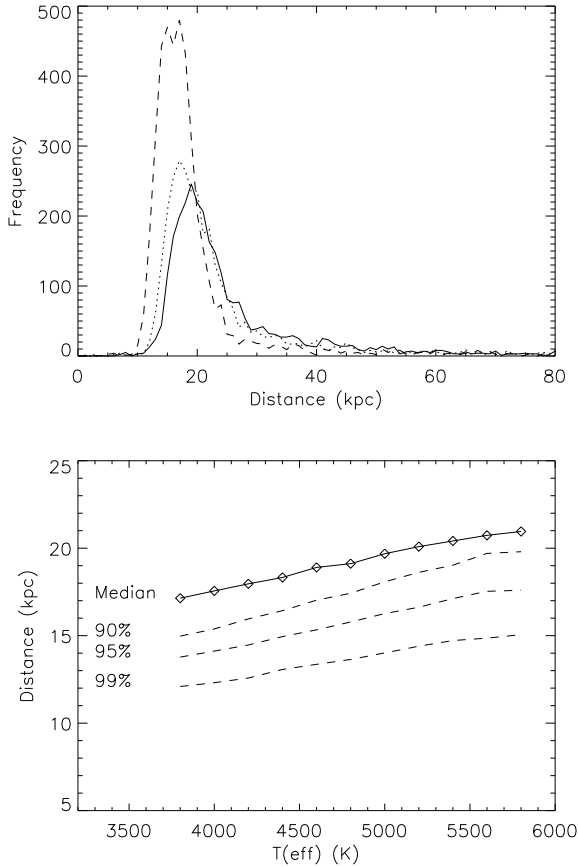}
 \caption{{\bf (Top:)} The probability distributions of the distance to 4U 1957+115 based on the 10000 Monte Carlo simulated synthetic SEDs fitted with models including a 5800K black-body component representing a solar type secondary star (solid line), a slightly evolved secondary (K2, 5000K - dotted line ) and a heavily evolved secondary (3800K - dashed line). Note that in $>$60\% of cases (no detection of the secondary) the distance is $>$80kpc. {\bf (Bottom:)} The median distance (in case when a secondary is detected) and the 90\%, 95\% and 99\% lower limits for the distance to 4U 1957+115 as a function of secondary spectral type based on 11x10000 Monte Carlo simulations. Note that the median distance refers to the median of cases where the secondary has a significant contribution (i.e. distance $<$50kpc).}
\end{figure}

\subsection{The Secondary star}

Given the reasonable fits to the SED by the accretion disc emission model alone, it is evident that the emission from the secondary star must be marginal and there is no clear visual detection of the secondary in our data. In order to obtain upper limits for the emission from
the secondary (and thus lower limit for the source distance) we have employed two approaches. Firstly, we have used our 
basic accretion disc model fits (as described above, no secondary included) and measured the scatter (standard deviation) in our K-band model flux from the model fits to our synthetic, i.e. Monte Carlo, datasets. We then use this scatter to derive a 3$\sigma$ upper limit for the K-band flux from the secondary. The resulting value is  9.0 $\times$ 10$^{-19}$ ergs cm$^{-2}$ sec$^{-1}$ \AA$^{-1}$ , which corresponds to a minimum distance of 15 kpc (assuming a solar type secondary) or 13 kpc (assuming an evolved 5000K secondary).

As a second method, we have added a an extra 5800K black-body component to our model and fitted the SED with this
more complex model and then carried out Monte Carlo simulations based on the fit in order to derive a probability distribution for the 
flux originating from the secondary. This was then transformed into a probability distribution of distance by adopting a solar absolute K magnitude of 3.33 (Worthey 1994). The resulting probability distribution peaks at 19 kpc (fig 3., top). We can also measure the 1\% lower percentile at 15 kpc. Naturally we cannot give any upper limit for the distance, given there is no firm detection of the secondary in the SED. Actually, in about $>$60\% of cases in our Monte Carlo simulations there is no significant flux from the secondary. The source distance probability distribution tail corresponding to these $>$60\% is off-scale (i.e $>$ 80 kpc) in Fig. 3.     

The results from our two different approaches to estimate the minimum distance of 4U 1957+115 seem to agree with each other
($\sim$15 kpc). Naturally these depend on the assumption that the secondary is of solar spectral and luminosity class. However, we know from the studies of secondary stars in cataclysmic variables of similar orbital periods, that the Roche lobe filling secondaries are typically somewhat evolved (Beuermann  et al. 1998, Knigge 2006). We estimate, using the empirical spectral-type vs. orbital period distribution of Knigge (2006), that the characteristic spectral type for the secondary is K2. i.e. the
effective temperature would be around 5000K. Lowering the black-body temperature to 5000K would have an approximate effect of  a 30\% drop in the K-band flux (retaining the same radius for the secondary and approximating it as a black-body). This would decrease the minimum distance from 15 kpc by about 15\% to 13 kpc. To further demonstrate this, we have carried out the previously described Monte Carlo analysis using a range of (fixed) secondary
effective temperatures from 3800K (M0) to 5800K (G2). The resulting distance distributions for 3800K, 5000K and 5800K cases are shown in the top panel of Fig. 3, whilst the median distance (of cases where the secondary is detected), together with the 90, 95 and 99\% lower limits as a function of secondary effective temperature are shown in the bottom plot of Fig. 3. Finally, given these lower limits to the distance, it is worth noting that in most cases of our Monte Carlo simulations the secondary is not detected and that the 1\% lower percentile limits for the distance are 15 kpc and 14 kpc for the solar type and K2 secondaries respectively. The resulting distance probability distribution (Fig. 3, top) is also highly skewed, setting very strict lower limits for the distance.  

We have also carried out a confidence region analysis in $\Delta \chi^2$ space in order to visualize the limits and correlations between the fit parameters. The results of this exercise are shown in Fig. 4 and discussed further in the Discussion section.

\subsection{The effects of disc irradiation and extinction}

In order to further examine the validity of our modelling described above, we have also carried out SED modelling using 
an approach introduced by Hynes et al. (2002). Their model consists basically of two power-law distributions for
the disc radial temperature i.e. one for the viscous disc ($\alpha$=-3/4) and another for the irradiated disc ($\alpha$=-3/7) 
(see Hynes et al. 2002 for details). The free parameters are the outer disc temperatures for
the ``viscous power-law" and the ``irradiated power-law" together with a normalisation term. The main difference to our
modelling is that the power-law indices are fixed. As a result we find that the ``irradiated power-law" component will
be attenuated in the fitting process and we are left with only the viscous (i.e. $\alpha$=-3/4) power-law, thus producing
the same best fitting model as in our original modelling. Exactly the same happens if we add a secondary component
to the Hynes et al. (2002) model. In practice, the outer disc temperature of the ``irradiated power-law" ($\alpha$=-3/7)
will be forced towards zero by the fitting process. 

Extinction is a free parameter in our fits and both our model and the Hynes et al. model favor $E_{B-V}$=0.0. This is somewhat 
surprising, given that both the X-ray spectral fits (Nowak et al. 2008) and the NASA/IPAC extinction estimates (based on Schlegel, Finkbeiner \& Davis (1998) dust maps) indicate $E_{B-V}$=0.2 and 0.175 respectively. However, our confidence region analysis (Fig. 4) shows that even if  $E_{B-V}$=0.0 yields the best fits, 
$E_{B-V}$=0.2 solutions still lie within the 90\% confidence region for our models. We have also tried forcing $E_{B-V}$=0.175
in our fits, but this makes the fits considerably worse ($\chi^2$ becomes 7.86 instead of 3.92, 2 degrees of freedom). If we
allow for much steeper radial temperature distributions, we can achieve equally good fits with $E_{B-V}$=0.175, but
this requires a disc inner edge temperature of 3.6$\times 10^{5}$K combined with a very steep radial temperature distribution of $\alpha$=-1.5. In this case we obtain $\chi^2$ = 3.97 (2 dof). However, there are no clear physical grounds for such an extreme$\alpha$ value.

\begin{figure*}
\includegraphics[scale=0.75,angle=90]{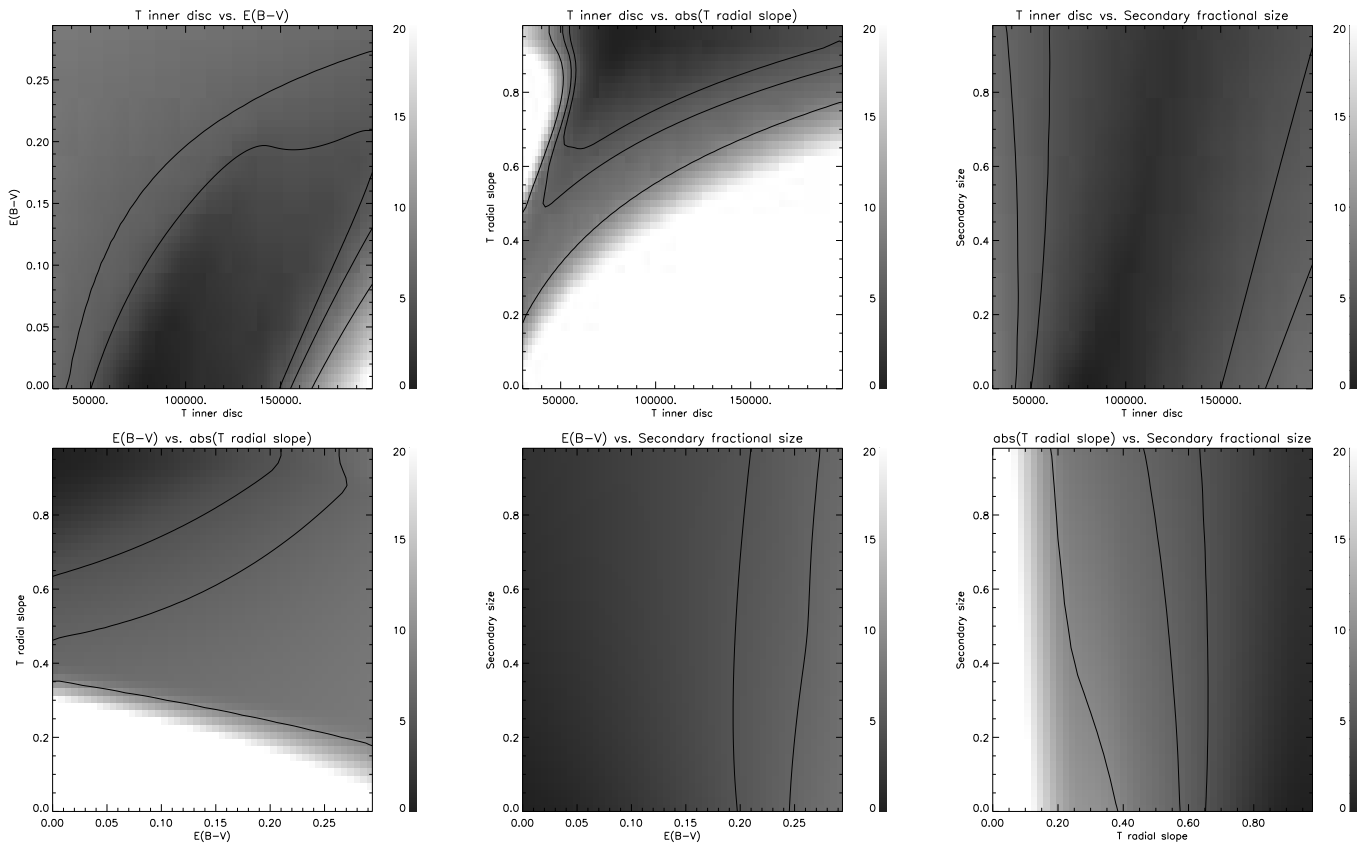}
\caption{The $\Delta \chi^2$ distribution maps and confidence contours for various parameter pairs. The darkest areas correspond to the minimum $\chi^2$ regions. We also show 90\%, 95\% and 99\% confidence contours obtainable from our SED fits. The results are obtained from a 100x50x50x50 grid of models covering the parameter ranges indicated in the plots.}
\end{figure*}

\section{Time series photometry}

In addition to obtaining the multicolour optical NIR magnitudes, we have also monitored the system in J, with interleaved V measurements at regular intervals (see Table 1). The motivation for this was to study whether the same 9.33 h modulation that appeared to have no colour dependence in UBVRI (HMD99) would continue in the NIR.

We have folded our J-band photometry, together with the quasi-simultaneous V-band data, over the 9.33 h cycle (presumed to be the orbital period) in Fig. 1. The J-band full amplitude is $\sim$ 0.25mag, similar to that observed by Thorstensen (1987) in the optical, but much lower than observed by HMD99. However, the light curve shape, with different gradients for
the decline and rise phases, is more similar to that detected by HMD99 than the sinusoidal light curve seen by Thorstensen (1987) or Bayless et al. (2011). It is clear that the light curve profile shows long-term evolution, possibly related to a precessional cycle of the accretion disc. Our V-band data points seem to follow the J-band modulation. This agrees with HMD99, where all the colours (UBVRI) seemed to show very similar modulation over the orbital period.  

\section{Discussion}

Several papers (Thorstensen (1987), HMD99, Russell et al. (2010), Bayless et al. (2011), Mason et al. (2012)) have reported     
results of optical studies of the system with rather varied conclusions about the origin of the optical variability. In summary, it
appears that the amplitude of orbital variability can range from zero to 0.6 mag full amplitude. In order to accommodate such large 
changes, the origin of the emission must be flexible in nature. Since 4U 1957+115 is a persistent X-ray source, such changes cannot
be produced by a model where the sole driving mechanism for the optical variability is the X-ray heating of the secondary star, as suggested by Bayless et al. (2010). This point is further underlined by our optical-NIR SED, the modelling of which requires a very significant temperature gradient in the disc. Such temperature gradient cannot be achieved on the surface of an X-ray heated
secondary star. In addition, the almost identical variability amplitude seen in all the light curves, from UV to NIR (HMD99) and our quasi-simultaneous V and J-band light curves presented in this study, also point to the accretion disc as a source of most of the optical and NIR emission. 

The RXTE ASM light curve of 4U 1957+115 has shown evidence for $\sim$ 117d superorbital period (Nowak \& Wilms, 1999). It is thus feasible that the different optical light curve shapes observed by different authors are a consequence of observing the system at different superorbital phases. If the source indeed harbours a black hole, the accretion disc would be susceptible to tidal instabilities due to the extreme mass ratio, and precession of the disc would follow (Whitehurst \& King 1991). Similar changes in accretion disc structure, as evidenced by changing optical light curve shapes,  are fairly common in LMXBs and have been observed in several systems such as Her X-1 (Gerend \& Boynton 1976), UW CrB (Hakala et al. 1998, 2009), 4U1916-05 (Callanan 1993, Homer et al. 2001), AC 211 (Auriere et al. 1989, Ilovaisky 1989), EXO 0748-676 (Pearson et al., 2006) and GR Mus (Cornelisse et al. 2013) to name just a few.   

It is rather peculiar that the SED modelling seems to require a rather steep temperature distribution with $\alpha = -0.75$. Even if such a distribution is predicted for purely viscous heating of an accretion disc (Shakura \& Sunyaev, 1973), a much flatter distribution (i.e. $\alpha = -0.5$) would be expected for a persistent X-ray source irradiating the disc (King 1998). However, if the irradiating X-ray source lies above the orbital (and disc) plane and the disc is mostly confined to the orbital plane, an extra $1 / r$ term is required 
to account for the disc projected area as seen by the X-ray source. Thus the X-ray heating per disc unit surface area would follow $F_X  \propto 1/r^3$ and assuming LTE ($F_X \propto T_{e}^4$), the disc surface temperature $T_{e}$ would follow $T_{e} \propto r^{-3/4}$ (Hynes, 2010). This means that for a realistic disc (i.e. a disc that is partly warped out of the orbital plane and heated by an X-ray source above
the disc) we expect the X-ray heating  to follow $F_X  \propto 1/r^n$, where $n=2-3$ (i.e. $T_{e} \propto r^{\alpha}$, where $\alpha=[-0.75,-0.5]$). We limited $\alpha$ to be $>$-0.75 in our fits to keep the results within physical bounds, but still found that the SED fits were typically pegged to this hard lower limit. It is however possible, that a convex inner disc structure or warping in the inner disc could shield the outer disc from irradiation, resulting in a much steeper temperature gradient than theoretically expected (Dubus et al., 1999). Similarly, if only the inner disc is surrounded by an X-ray scattering hot accretion disc corona (ADC), the scattered X-rays, depending on the ADC optical depth and size, will preferentially heat the part of the disc within the ADC radius. In fact, our grid-based exploration of the parameter space (Fig. 4) covered an $\alpha$Ê-range of [-1.0,0.0] and showed that the best fits could be obtained with the steepest temperature distributions, thereby pointing towards extra heating in the inner disc. Finally we would like to point out that, apart from the grid-based computations (Fig. 4) where a wide parameter space was systematically explored, the maximum area of the secondary star in the fits was limited to 1/3 of the projected area of the accretion disc.
This is a reasonable assumption since, for instance, taking a likely value of $q$=0.1 (assuming a BH system) the Roche lobe area of the secondary becomes about 10-15\% of the accretion disc area. Now, assuming a flat disc and maximum inclination of 70$^{o}$ 
(no eclipses are observed) will bring this fraction up to 30-35\%. Without such limitation, the best fit model would include a secondary 10-50x the size of the accretion disc together with a very steep hot disc continuum, which is clearly not physical.   

It is interesting to note that taking our accretion disc model of the SED and extrapolating the temperature distribution inwards yields a maximum disc temperature of 0.8-0.9 keV at the Schwarzschild radius for a 15 M$_\odot$ black hole. This is in agreement with
the maximum disc temperatures seen in black hole binaries. Nowak et al. (2008) list disc black-body (XSPEC model {\it diskbb}) 
temperatures in the range 1.2-1.8 keV for 4U 1957+115. Given the uncertainties involved in our extrapolation, these values are
in agreement with our model. It is thus feasible that all the emission ranging from soft X-rays to NIR can be explained largely by a single multi-temperature accretion disc model with the standard radial disc temperature profile proportional to $r^{-3/4}$. 
Similar results have been obtained by Russell et al. (2011), who find that the broadband (radio to X-ray) SED can be modelled
with thermal emission from a multi temperature accretion disc alone. The common IR excess, attributed to synchrotron emission 
seen in many black hole XRBs in outburst (Russell et al. 2006), is not detected. The likely reason is that the source is in a persistent soft state, with no radio jet emission (Russell et al. 2011).
This is in agreement with our preferred model, where the optical/NIR modulation is due to the changing projected area of
either a warped or flared accretion disc over the orbital period and there is no significant contribution to the optical/NIR emission
from the secondary star. 

Our 99\% lower limit for the distance of the source is $\sim$15 kpc. This, together with the galactic latitude of 9.3 degrees, places the system 2.5 kpc above the galactic plane and at about 12 kpc from the galactic centre, making it a distant galactic halo object. Adopting a distance of 18 kpc (peak in the probability distribution, Fig 3.) and taking the 0.7-8 keV X-ray flux of 1.0 $10^{-9}$ erg cm$^{-2}$ sec$^{-1}$ from Nowak et al. (2012) would then imply an X-ray luminosity L$_X$ = 3.8 10$^{37}$ erg sec$^{-1}$ which is about 15\% of the Eddington limit for a 1.4M$_{\odot}$ neutron star (or less than 2.5\% for a 16 M$_{\odot}$ black hole). Russell et al. (2010) examine the X-ray-optical luminosity ratio of 4U 1957+115 in comparison with a variety of X-ray binaries. They conclude that if the distance is more than 20 kpc, the source is likely to contain a black hole (or possibly a NS atoll source if the distance is of the order of 20 kpc). Our data is thus inconclusive regarding the nature of the compact object, even if in about 80\% of cases in our Monte Carlo study the distance is above $\sim$ 20 kpc.

There now appears to be a small population of short period (2-5 h) black hole candidates known in the galactic halo (Shaw et al. 2013), all of which are X-ray transients and most of which are substantially sub-Eddington for their (dynamically) determined masses. If the high mass of the compact object in 4U 1957+115 can be confirmed, the source would be the first persistent, and the longest period, BHC to join this population. It is worth noting  that based on its X-ray luminosity (and inferred accretion rate), 4U 1957+115 is likely to be a persistent source according to the disc instability models (Coriat et al. 2012).

\section{Conclusions}

We have modelled the BVRIJHK-SED of 4U 1957+115 with an accretion disc model that allows for different radial temperature
 profiles i.e. different disc heating mechanisms. Our results are in agreement with a viscously heated disc, but given the
 uncertainties involved in the expected irradiatively heated disc profile and the persistency of the X-ray source, we still believe 
 that the X-ray heating of the disc must play a major role.
    
Furthermore, modelling our optical and NIR photometry with a multi-temperature accretion disc model together with
a solar-type secondary yields a  99\% minimum distance of 15 (14) kpc to 4U 1957+115. The value in parentheses correspond to an evolved K2 secondary. 
This distance is broadly in agreement with the 16M$_\odot$ black hole models with spin parameter $a^* \sim$1 and spectral hardening factor of $h_d$ = 1.4 as proposed by Nowak et al. (2008). Our quasi-simultaneous J and V-band photometry confirms that the optical 
and NIR emission indeed come from a common location i.e. the accretion disc. This means that the time dependent/
long term changes in the optical and NIR orbital modulation are likely caused by changing orientation of the 
precessing accretion disc within the binary rest frame. Simultaneous optical and NIR observations, as well as a high resolution NIR spectroscopic search for a signature of the donor, would be highly desirable in order to build up a more coherent picture of the underlying spectral model and put even more stringent limits on the secondary star and the distance to 4U 1957+115. This is particularly important since our current set of data is only broadly consistent with the independent extinction estimates towards the object and we particularly need more simultaneous UV and optical data.

\section*{Acknowledgments}

Based on observations made with the Nordic Optical Telescope, operated by the Nordic Optical Telescope Scientific Association at the Observatorio del Roque de los Muchachos, La Palma, Spain, of the Instituto de Astrofisica de Canarias. This research has made use of the NASA/ IPAC Infrared Science Archive, which is operated by the Jet Propulsion Laboratory, California Institute of Technology, under contract with the National Aeronautics and Space Administration. We would like to thank the anonymous referee
for input that made the final paper significantly better.

\end{document}